# Bringing Microbiology to Light: Towards all-Optical Electrophysiology in Bacteria

Giuseppe Maria Paternò[1]*, Gaia Bondelli[1,2], Guglielmo Lanzani[1,2]

[1]Center for Nanoscience and Technology, Istituto Italiano di Tecnologia, Via Pascoli 70/3, 20133, Milano, Italy.
[2]Physics Department, Politecnico di Milano, Piazza Leonardo da Vinci 32, 20133 Milano, Italy.
*Correspondence to: Giuseppe.paterno@iit.it

**Abstract**

The observation of neuron-like behaviour in bacteria, such as the occurrence of electric spiking and extended bioelectric signalling, points to the role of membrane dynamics in prokaryotes. Electrophysiology of bacteria, however, has been overlooked for long time, due to the difficulties in monitoring bacterial bioelectric phenomena with those probing techniques that are commonly used for eukaryotes. Optical technologies can allow a paradigm shift in the field of electrophysiology of bacteria, as they would permit to elicit and monitor signalling rapidly, remotely and with high spatiotemporal precision. In this perspective, we discuss about the potentiality of light interrogation methods in microbiology, encouraging the development of all-optical electrophysiology of bacteria.

## Introduction

The current golden standard for electrophysiology is represented by the patch clamp technique, which permits to measure membrane electrical phenomena precisely and directly. However, patch clamp and, in general, electrical probing and stimulation techniques, require physical contact with the target cells and are usually time-consuming. On the other hand, optical technologies offer the possibility to target specific cells or even cell compartments at the same time with high spatiotemporal precision, remotely and in a non-invasive fashion. Therefore, driven by these motivations, many research efforts have been directed towards the development of non-invasive and high-throughput optical methods in electrophysiology.

Neuroscience has largely benefitted from optical technologies so far, as the efficient and reversible modulation of neuronal activity allows studying the function of specific brain circuits at unprecedented time/space scales, thus opening new avenues for the treatment of psychiatric and neurological diseases.[1] Optogenetics is the pioneering technique for eliciting and probing neuronal response by means of light.[2] This approach exploits gene therapy to express exogenous light-sensitive transmembrane proteins, enabling both on–off control of neuronal activity with cell-type specificity and recording of intercellular signalling via genetically-encoded voltage reporters.[3] Photopharmacology is an alternative method to confer light-sensitivity to cells.[4] In this case, selected light actuators (*i.e.* azobenzene tethered ligands) are directly attached to the specific ion channel[5], at the lipid membrane level,[6] or even to drugs to photomodulate their pharmacodynamics and pharmacokinetics.[4] Both optogenetics and photopharmacology, however, require either genetic modification or covalent attachment of light-actuators to the target, thus introducing a set of technical challenges. An alternative and relatively novel paradigm in opto-stimulation relies on the use of nanostructures and molecular actuators to achieve optical control over bioelectric signalling without the need to modify genetically or chemically the bio-target. Relevant examples include the use of inorganic and organic nanostructures and molecular systems to modulate neuronal activity, and to rescue vision.[7,8] In general, these methods possess lower specificity and temporal responses than optogenetics and photopharmacology, while ensuring comparable electrophysiological performances with lower invasiveness.[9]

Recent findings have demonstrated that membrane potential dynamics and extended bioelectric signalling occur also in bacteria.[10,11] This offers a chance for exploiting opto-stimulation methods to prokaryotes. The study of brain-like signal propagation in bacteria is of fundamental interest as a model system of neuronal network,[12] but it has also direct implication in bacterial behaviour, metabolism and adaptation to antibiotics, among others.[13,14] In these regards, the relatively new field of bacterial electrophysiology can benefit extensively from the utilisation of optical methods. Light holds the potential to overcome the existing limitations of the electrode-based techniques, mostly arising from the requirement to contact small and motile cells exhibiting thick cell walls.[15] Therefore, the development and engineering of new photoactive systems capable to interface effectively with bacteria and to trigger/probe signalling is highly desirable.

In this perspective, we aim at providing an outlook over the use of optical methods to control and probe biological signalling, with particular emphasis on the possible use of light technologies for the investigation of the bacterial electrophysiology. It is worth adding that while optogenetics and non-genetic optical methods are being applied to trigger and/or investigate signalling in bacteria, photopharmacology is traditionally utilised for the light-modulation of antibacterial activity of existing antibiotics, *i.e.* via the covalent attachment of azobenzene ligands to the target molecules (see for instance the works of the Feringa's group and collaborators[16–18]). Therefore, we will not include photopharmacology in our discussion, as this method is not directly employed for the photocontrol of bioelectric signalling in bacteria.

**Electrostimulation and probing methods in bacterial electrophysiology**

Electrostimulation and probing methods are currently being applied for studying the electrophysiology of bacteria. In these regards, Asally and collaborators have shown recently that membrane potential dynamics can be elicited via stimulation with exogenous electrical signal.[14] For this purpose, the authors designed a bespoke electrode-coated glass-bottom dish, which allowed electrostimulation while ensuring membrane potential probing with a fluorescence membrane-potential indicator. In particular, they showed that electrical stimulation causes hyperpolarization in unperturbed cells. Conversely, when cells were pre-exposed to UV light or antibiotics, the same electrical stimulation depolarizes cells instead of causing hyperpolarization. This suggested that external stimulation can be used as a tool for the discrimination between proliferative and non-proliferative bacteria. Passing to electrical probing, patch clamp is the technique of choice in neuroscience. Although it permits to measure directly and quantitatively the electrical potential it cannot be applied to living bacteria, due to their small sizes and cell wall. On the other hand, patch clamp can still provide some precious information on the role of ion channels in bacteria.[19,20] For example, Martinac et al. used patch clamp on giant spheroplasts to study native mechanosensitive channels in *Escherichia Coli*.[21]

The most important advantage of electrical techniques resides on the fact that they have been utilised largely in neuroscience for many years. Electrical stimuli can permit to trigger the potential dynamics with spatial precision, both in planktonic cells and in biofilms by using multi-electrode arrays.[22] However, optostimulation can surpass electro-based methods in some regards. Specifically, light can be used to target both specific cells and subsets of a bacterial community at the same time and with different colours, also via the use of patterned excitation. For instance, this can be a key advantage to study cell-to-cell signalling heterogeneity in biofilms, which accounts for antibiotic stress response[23] and the xenobiotic metabolism in the gut microbiota.[24] In addition, light stimuli can enable fast and localised excitation at the close proximity or even within the bacterial cell, permitting to potentially disclose and investigate fast (ms) signalling in bacteria. Finally, biomaterials science can permit to optimise the photoactuation process for a given bacterial species or, alternative, can allow engineering of actuators that are well-suited for a wide range of bacteria, in the view to utilise optostimulation also in complex communities. Passing to the membrane potential probing, our idea is that optical methods are intrinsically more advantageous than patch clamp for bacterial electrophysiology. This explains why fluorescent voltage indicators are already used extensively for this purpose. However, these systems do not allow a direct and quantitative measurement of the potential, as it happens in patch clamp. This can be a serious limitation for the development of the field.

## Optogenetics

In optogenetics, light sensitivity is conferred to cells via the incorporation of a protein-based light-actuator that is fully genetically encoded. In its classical form, optogenetics employs microbial opsins to take photocontrol over neuronal signalling. The first seminal study on the photocontrol of neuron spiking via the use of channelrhodopsin-2 was reported by Deisseroth and collaborators in 2005.[25] This report demonstrated for the first time that temporally precise and non-invasive control of neuronal activity can be achieved at faster timescales (milliseconds) than previous photostimulation methods, such as photopharmacology.[26] Since then, a plethora of genetically encoded light actuators and fluorescent voltage reporters have been developed.[1,3,27] These include various natural photoreceptors from animal, plants and microorganisms.

In regards to microbiology, optogenetics has been employed recently to achieve photocontrol over the swimming speed of bacteria,[28,29] a parameter that ultimately depends on the protonmotive force and membrane potential. On the other hand, genetically encoded voltage indicators (GEVIs) are largely utilised in bacterial electrophysiology.[13] In 2011, Cohen and collaborators reported for the first time on the occurrence of electrical spiking at 1 hertz in *E. coli* (**Fig. 1**), which was revealed by using a genetically encoded dye, namely proteorhodopsin optical proton sensor (PROPS).[30] In particular, the indicator showed a fluorescence blinking behaviour that was directly related to membrane electrical depolarization. This seminal study opened the field of bacterial electrophysiology, a research area that was usually poorly investigated due to small size of bacteria and a lack of sensitive tools. For instance, Bruni et *al*. employed a combined calcium-voltage sensor based on PROPS to show that electrical spiking is connected to mechanosensation.[31] More recently, the same indicator has been used to prove that the bactericidal action of aminoglycosides, such as kanamycin, arises from the dysregulated membrane potential. In the absence of transmembrane voltage, aminoglycosides are taken into cells and exert bacteriostatic effects by inhibiting translation.[32]

To summarise, although optogenetics would allow stimulation and monitoring of membrane potential dynamics at unprecedented length scales, this technique would unavoidably introduce a set of challenges, *i.e.* the poor applicability of the genetic transformation to certain bacterial species and strains,[13] as well as the large use of UV light that already interfere with membrane potential.[33] This might possibly represents a serious limiting factor for the widespread use of optogenetics for bacterial electrophysiology.

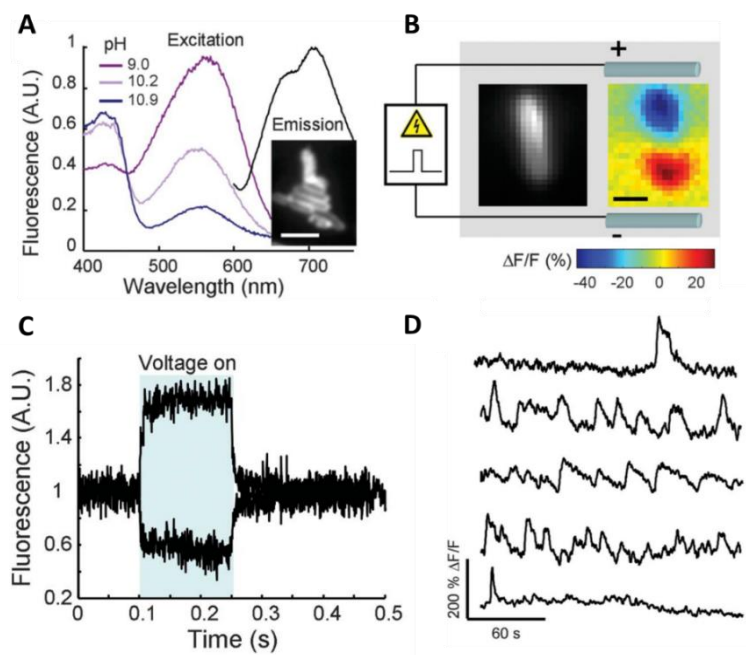

**Figure 1.** PROPS dye is used as fluorescent indicator to monitor membrane potential dynamics. (A) Excitation/emission spectra of PROPS showed pH-dependent fluorescence, making it a promising candidate voltage sensor. (B) Spatially-resolved

change in fluorescence in a cell subjected to induced transmembrane voltage (ITV). (C) Fluorescence response to positive and negative steps in membrane potential. ITV was used to calibrate fluorescence *vs.* voltage. (D) Membrane potential dynamics of five single cells. Adapted from reference.[30] Copyright 2011, American Association for the Advancement of Science.

## Non-genetic optostimulation

In this strategy, the photoactuator and/or the probe are neither genetically encoded nor chemically linked to the bio-target, as it happens for optogenetics and photopharmacology, respectively. Thus the abiotic/biotic interface is constituted on the basis of non-covalent affinity or physical contact with the cells. In the last couple of decades, this field has seen a steep rise mostly due to the development of new nanostructured materials and molecular actuators. These systems permit to localise the excitation with a relatively high spatial confinement owing to their reduced dimensionalities, as well as offering the possibility to tune their action spectrum and temporal resolution via top-down processes and chemical synthesis.[8]

Photomodulation usually occurs via three main mechanisms, namely: photothermal, photocapacitive and photoelectrochemical phenomena due to charge generation, and photomechanical. Both the photo excitation properties of the materials and the interface structure (*i.e.* diffuse or planar interface) dictate which bio-stimulation mechanism is at work. Detailed information about the most employed materials and interfaces used for non-genetic optostimulation, as well as the relative stimulation mechanisms, are contained in the following recent reviews,[7,8,34] and are briefly summarised in **Fig. 2**. Conversely, non-genetic photomechanical stimulation has been introduced for eliciting signalling in neurons only recently. In particular, it has been proposed a new amphiphilic azobenzene that dwells in the plasma membrane in a non-covalent manner, giving rise to a light-induced transient hyperpolarization followed by a delayed depolarization that triggers action potential both *in vitro* and *in vivo*.[9,35–37] The stimulation mechanism resides in the *trans*→*cis* photoreaction of azobenzenes: molecular dynamics simulations show that in dark the *trans* isomer can undergo dimerization causing a thinning of the membrane and an increase of its electrical capacitance, while illumination generates the formation of a stable population of *cis* isomers and, thus, the disruption of the dimers leading to a restoration of membrane thickness and capacitance.

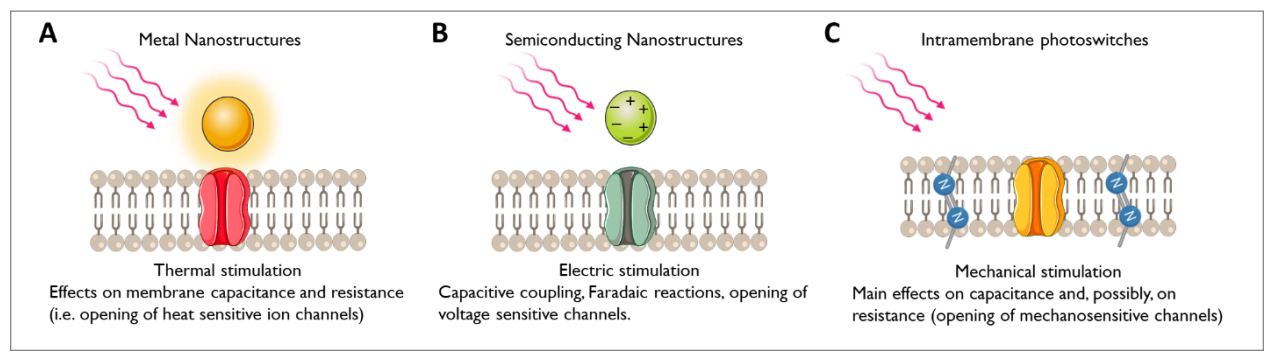

**Figure 2**. Cartoon illustrating the three main approaches and materials employed for non-genetic optostimulation. (A) Thermal optostimulation usually relies on the use of metal nanostructures,[38,39] owing to the efficient and rapid heat dissipation upon photoexcitation of charge carriers in these systems. Local heating can have a large effect on membrane capacitance and resistance,[40] for instance via the opening of heat sensitive ion channels. (B) In semiconducting nanostructures, photoexcitation leads to the generation of electrons and holes. The photoinduced electric field can cause either the opening of voltage sensitive channels and/or to the occurrence of faradaic reactions between the actuator and the cell. (C) Intramembrane conformational photoactuators, such as azobenzene derivatives, can be used to elicit a membrane potential dynamics without directly affecting ion channels or local temperature.[9,35] In this case, the conformational change mostly affects membrane thickness and its electrical capacitance.[36]

Currently, non-genetic optostimulation methods are largely applied to neuroscience, for instance for eliciting signalling response in neurons[8] and for vision restoration,[41,42] and for developing new bioelectronic plafforms.[43,44] On the other hand, this approach holds a great potential for delivering transient (from milliseconds to picoseconds), localised and controllable perturbations to bacterial cells and communities, without the need to use complex wiring or specific stimulation devices. These advantages can make light stimulation a valid and powerful alternative to electrical interrogation methods,[14,45,46] *i.e.* permitting to study fast microbial signal transduction in response to transient perturbations. The first evidence of the potentiality of light stimulation methods in microbiology have been demonstrated in a recent paper.[47] In particular, Gao *et al.* have reported on the use of silicon micro/nanostructures to trigger signalling in planktonic bacteria and communities via photothermal stimulation. Interestingly, this effect permits to elicit a previously unidentified form of rapid, photothermal gradient–dependent, intercellular calcium signalling within the biofilm (**Fig. 3**).

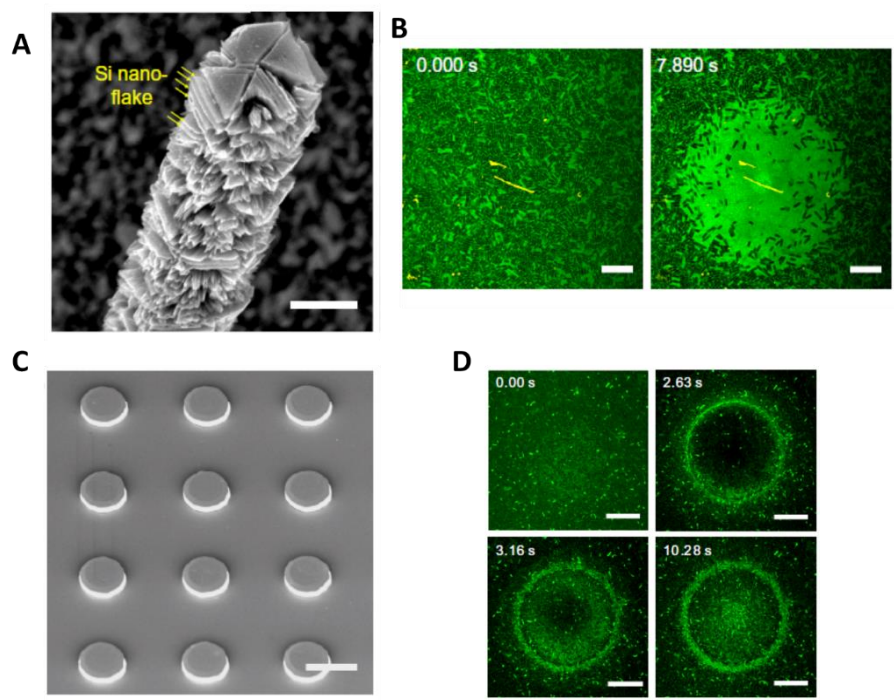

**Figure 3.** (A) Scanning electron microscopy (SEM) image of a mesostructured Si nanowire synthesized via chemical vapour deposition (scale bar = 0.5 µm). (B) Confocal microscope images of light-induced $Ca^{2+}$ elevation in a *B. subtilis* biofilm triggered by silicon nanowires, before (left) and after (right) stimulation (592 nm, ~22.3 mW, 1 ms, scale bar = 10 µm) as measured with Fluo-4 AM. (C) SEM image of 2D Si discs (~2 µm thickness) made from top-down fabrications (scale bar = 10 µm). (D) Confocal microscope time series showing the evolution of $Ca^{2+}$ distribution over time. The authors have found that the calcium signalling pattern depends on the microplate geometry. Adapted from reference[47]. This is an open-access article distributed under the terms of the Creative Commons Attribution-NonCommercial license.

However, the simple extension of the optostimulation tools used in neuroscience to prokaryotes might be still insufficient for a rapid advancement of the field. This is largely due to the differences between eukaryotic and prokaryotic electrophysiology, stemming essentially from the relatively small sizes and presence of cell wall in bacteria. Details about the electrophysiological dynamics in prokaryotes are discussed comprehensively in two review articles.[13,15] Briefly, the smaller surface area of bacteria than mammalian cells (6 μm² vs. 1600 μm²) leads to a lower membrane capacitance in prokaryotes ($V_m = q/C$, where $V_m$ is membrane potential, $q$ is charge amount and $C$ is the membrane capacitance) meaning that a smaller difference in capacitance gives rise to higher membrane potential changes. Furthermore, small sizes in bacteria translates into small number of cytoplasmic ions, thus implying that opening of ion channels and decrease of membrane resistance will rapidly lead to a depletion of the less abundant ions in few seconds.[13,19] Moreover, another important difference lies in the structure of the cell envelope, with gram-positive and gram-negative cells exhibiting a relatively high density of negatively charged molecules at their cell walls (lipopolysaccharides for gram-negative and lipoteichoic and teichoic acids and surface proteins for gram-positive bacteria). Altogether, these profound differences suggest some general criteria for the design of photoactuators *ad hoc* for bacterial non-genetic optostimulation. Given that membrane capacitance and resistance can be a sensitive targets for stimulation, an ideal photoactuator should be able to directly focus on these parameters. In neurons, this is usually accomplished by means of thermal increase of the local temperature via the excitation of metal nanostructures and capacitive coupling with semiconducting materials (see **Fig. 2**), with both methods acting indirectly on the ionic species and electric field in the vicinity of the membrane. In prokaryotes, the presence of cell wall and the small number of extracellular ions would in principle decrease the impact of these approaches. Alternatively, we reckon that molecular materials that directly partition in the plasma membrane would be more effective in these regards. These could act on the geometrical characteristics of the membrane and/or on its polarizability (*i.e.* donor-acceptor molecules), thus leading to a direct photomodulation of the membrane capacitance or to the opening of mechanosensitive or voltage-gated channels (**Fig. 4**).[19,21] Last but not least, chemical synthesis usually enables fine tailoring over molecular characteristics. This can allow enhancing the affinity of the actuator for bacterial cells and the membrane environment, for instance via the judicious use of bio-affine charged groups capable to interact with the negatively charges of the cell wall, and appropriate lipophilic units that drive partitioning in the membrane interior.

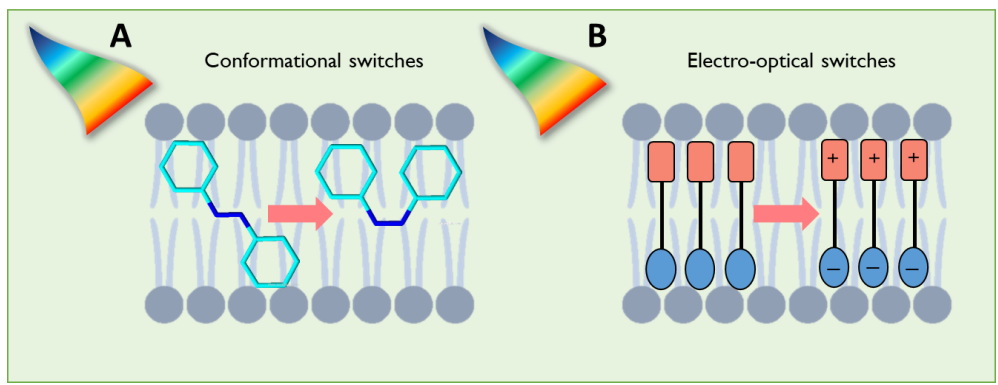

**Figure 4.** Two possible examples of intramembrane molecular switches that could permit to elicit efficiently the bacterial membrane potential dynamics. These systems would directly modulate the membrane capacitance via geometrical perturbation of the bilayer characteristics (i.e. thickness) by using conformational switches (A), or via a photoinduced change of its electrical polarizability by employing intramolecular electron donor-acceptor molecules (B).

Together with the introduction of efficient light-actuators, the exploitation and development of non-genetic fluorescent indicators for the non-invasive monitoring of cell membrane potential can permit to advance towards all-optical bacterial electrophysiology. Currently, the most used non-genetic membrane potential indicators are Nernstian dyes and membrane-bound dye. The functioning mechanism of Nernstian dyes rely on the different distribution across the membrane of cationic molecules, whose partitioning is regulated by the Nernst potential. Therefore, it is possible determining the transmembrane potential by measuring the fluorescence ratio of the dye inside and outside the cell. The most used Nernstian dye in bacteria electrophysiology are tetramethylrhodamine, methyl ester (TMRM) and bis-(1,3-dibutylbarbituric dcid) trimethine oxonol (DiBAC$_4$(3)). These systems, however, require careful calibration for a quantitative estimation of the membrane potential.[48] In addition, their relatively high permeability hampers the investigation of fast dynamics.[13] Conversely, membrane-bound dyes are well suited for monitoring fast membrane potential dynamics, even though they exhibit a weaker signal-to-noise ratio than Nernstian dyes. The most used membrane-bound dye in bacteria electrophysiology is aminonaphthylethenylpyridinium (ANEP).[49] Membrane-bound dyes shift their excitation spectra according to the membrane potential. The fast developments in the field of membrane potential indicators suggest that optical methods in electrophysiology will be more routinely used in the future. However, to compete effectively with patch clamp technique, these indicators should allow direct measurement of the membrane potential without the need of careful calibration procedures. To this end, a paradigm shift in the recording mechanism is highly desirable. A possible strategy for obtaining absolute information is to translate the spectral data into the time-domain, such as by using fluorescence-lifetime imaging microscopy (FLIM). For instance, the use of indicators that changes their excited state lifetime upon potential variation has been demonstrated as a powerful approach to measure absolute voltage in eukaryotes,[50–52] while this method has never been applied to prokaryotes.

## Conclusions and outlooks

In this perspective, we have summarized the most popular approaches employed to confer light sensitivity to cells, allowing both photomodulation and monitoring of cells signalling. To date, optical modulation and interrogation techniques have been largely applied in neuroscience. These approaches have allowed gaining insights into extended neuronal signalling and specific brain functions at unprecedented time/length scales. Interestingly, recent findings have indicated that bacteria possess neuron-like behaviour, such as electric spiking and extended bioelectric signalling, opening the way to electrophysiological studies in bacteria. In principle, bacterial electrophysiology can immensely benefit from optical methods, as light can surpass electrostimulation/probing in many regards, thus allowing a paradigm shift analogous to the introduction of optostimulation in neuroscience. To achieve this, the actuation process needs to be tailored specifically for prokaryotes, while optical probing should permit to access the membrane potential directly and quantitatively. Although this field is at its infancy, we reckon that interdisciplinary research efforts encompassing microbiology, biomaterials science, optical spectroscopy and imaging, will permit to reach the critical mass necessary to develop the tools and knowledge for bacterial all-optical electrophysiology.

## Author contributions

All authors contributed to manuscript drafting and revising, and figure creation. The final version of the article was reviewed and approved by all authors. This article has been accepted for publication in the journal "Bioelectricity".

## Conflict of interest

No competing financial interests exist.

## Funding information

We acknowledge Fondazione Cariplo for the financial support (grant n° 2018-0979 and grant n° 2018-0505).